\documentclass[letterpaper,conference]{IEEEtran}
\IEEEoverridecommandlockouts
\usepackage{paper-2026-dcoss-iot-ac-dpf}
\usepackage[anonymous=false]{myauthors}

\begin{document}

\title{Flexible \acl{DPF} for the \acl{IoT} via \acl{AC}}

\maketitle


\begin{abstract}
State estimation from uncertain, distributed observations is central in many cyber-physical applications.
While \ac{DPF} algorithms address non-linear and non-Gaussian estimations in distributed settings,
most solutions remain tied to specific architectures and communication assumptions,
limiting adaptability in open, heterogeneous deployments%
---most notably, the \ac{IoT}.
In this paper,
we propose a field-based formulation of \acl{DPF} grounded in \ac{AC}.
By expressing estimation and information dissemination as computational fields,
our approach decouples the core filtering logic from coordination and data-flow strategies.
This enables systematic customisation of key design dimensions, including
fusion-center placement and resilience,
aggregated measurement functions,
as well as 
the type and scope of information propagation.
Through a set of in-silico experiments,
we show how diverse \ac{DPF} configurations can be derived within a unified framework, highlighting
trade-offs among accuracy,
communication cost,
and
robustness.
Overall,
the proposed approach positions \ac{AC} as an effective abstraction layer
for engineering adaptable \ac{DPF} solutions in open \ac{IoT} environments.
\end{abstract}


\acresetall

\section{Introduction}

Estimating the state of a dynamic system from uncertain and partial observations
is a foundational problem in a wide range of application domains~\cite{HlinkaHD13},
including
target tracking,
monitoring,
and
cyber-physical control.
In many realistic scenarios,
the system of interest exhibits non-linear dynamics and non-Gaussian uncertainty,
making classical linear estimation techniques inadequate.
Particle Filters (PF)~\cite{DBLP:journals/tsp/ArulampalamMGC02} have emerged as a principled and widely adopted approach to address such settings,
providing a Monte Carlo approximation 
of sequential Bayesian estimation 
that can flexibly accommodate complex system and measurement models~\cite{HlinkaHD13}.

In \ac{IoT} and large-scale distributed systems,
observations are typically collected by multiple spatially distributed entities,
each endowed with limited sensing, computation, and communication capabilities.
This has motivated extensive research on \ac{DPF}~\cite{HlinkaHD13},
where the estimation process is decentralised across a network of cooperating nodes.
Over the years,
a rich ecosystem of \ac{DPF} algorithms has been developed,
spanning
fusion-center-based approaches,
leader-agent schemes,
consensus-based methods,
and
hybrid dissemination strategies.
These approaches explore different trade-offs among
estimation accuracy,
communication cost,
robustness,
and
scalability,
and have been successfully applied to a variety of scenarios.

Despite many solutions, 
state-of-the-art \ac{DPF} approaches achieve flexibility through specialised designs 
tightly coupled to specific architectural, communication, and modelling assumptions.
For instance,
the role of fusion centers,
the structure of information propagation,
the representation of beliefs,
and
the nature of measurement models
are often hard-coded into the algorithmic formulation.
As a result, 
adapting a given \ac{DPF} solution to changes in sensors, topology, reliability, 
or mobility often requires non-trivial redesign rather than simple reconfiguration.
This limits the suitability of existing \ac{DPF} approaches for open and heterogeneous \ac{IoT} systems,
where assumptions on deployment density, sensor quality, and connectivity are inherently weak and subject to change.

To address these challenges,
we turn to \ac{AC}~\cite{DBLP:journals/computer/BealPV15},
a macro-programming~\cite{DBLP:journals/csur/Casadei23} paradigm designed to express collective behaviour in distributed systems
independently of low-level coordination details.
\ac{AC} promotes a global, aggregate view of computation,
where programs describe how information should flow and be transformed across a population of devices,
rather than prescribing device-level protocols.
This perspective fits \ac{IoT} scenarios, 
abstracting over failures, asynchrony, 
and dynamic topologies while enabling concise resilient behaviours.

At the formal core of \ac{AC} lies the \ac{FC}~\cite{DBLP:journals/tocl/AudritoVDPB19},
a functional model for programming \emph{(computational) fields}~\cite{DBLP:journals/pervasive/MameiZL04},
i.e., distributed data structures mapping devices or spatial locations to values.
\ac{FC} provides language constructs to
evolve fields over time,
to exchange information among neighbouring devices,
and
to restrict computations to dynamically determined regions of the network.
Crucially,
\ac{FC} offers equivalent local and global semantics:
the same program can be interpreted both as a description of individual device behaviour
and
as a specification of the resulting collective dynamics.
This duality enables reasoning at the aggregate level while remaining implementable on fully decentralised systems.

In this paper,
we propose a field-based approach to \ac{DPF}
that integrates \ac{AC} and \ac{FC} as a unifying abstraction layer for distributed state estimation.
Rather than introducing a new \ac{DPF} algorithm,
our contribution lies in reframing \ac{DPF} as a family of configurable field computations,
where key design dimensions become explicit and composable.
These include
\begin{inlinelist}
    \item the placement and resilience of fusion roles,
    \item the definition of measurement functions
    (possibly aggregated across neighbourhoods),
    \item the form of information exchanged among nodes,
    and
    \item the spatial and topological scope of dissemination.
\end{inlinelist}
By expressing these aspects as computational fields,
the proposed approach supports
systematic adaptation to heterogeneous sensors,
uneven spatial observability,
node mobility,
and
failures,
without entangling such concerns with the core estimation logic.


We validate the proposed framework through
a set of in-silico experiments exploring multiple \ac{DPF} configurations
derived from the same field-based specification.
These include fusion-center-based schemes realised via leader election for fault tolerance and
fully decentralised solutions relying on local particle filters with aggregated measurement functions.
Overall,
our results show that a field-based formulation enables a flexible and principled design space for \ac{DPF},
well aligned with the requirements of open \ac{IoT} environments.

The 
 paper is structured as follows.
\Cref{sec:background} recalls the literature on \acs{PF}, \ac{DPF}, and \ac{AC},
and summarises the most relevant related works.
\Cref{sec:model} presents our modelling of the \ac{DPF} problem from an \ac{AC}, field-based perspective.
\Cref{sec:use-case} discusses the exemplary case of target tracking.
\Cref{sec:experiments} describes the experimental setup and discusses the obtained results.
Finally,
\cref{sec:conclusion} concludes the paper and outlines promising directions for future research.

\section{Background}
\label{sec:background}

\subsection{The Filtering Problem via \acf{PF}}
\label{sec:background-pf}

Sequential state estimation concerns
the reconstruction of the latent evolution of a dynamical system from partial and noisy observations.
In the context of \ac{IoT}, 
 the system state typically represents physical, environmental, or logical quantities of interest
-- such as position, speed, load, occupancy, or anomaly indicators --
that evolve over time and are indirectly observed by spatially distributed observer devices.

\paragraph{Underlying system model}

Let $\vec{x}_t \in \mathbb{R}^n$ denote the system state at discrete time $t$,
and let $\vec{y}_t \in \mathbb{R}^m$ denote the corresponding observations.
The system is commonly modelled as a state-space process
\begin{align}
\vec{x}_t &= f(\vec{x}_{t-1}, \vec{u}_t), \\
\vec{y}_t &= h(\vec{x}_t, \vec{v}_t),
\end{align}
where $f(\cdot)$ and $h(\cdot)$ are the (possibly non-linear) transition and observation models,
and $\vec{u}_t, \vec{v}_t$ represent process and measurement noise, respectively.

There,
the transition model $f(\cdot)$ describes the current system state ($\vec{x}_t$)
as a function of the previous state ($\vec{x}_{t-1}$) possibly subject to \emph{process} noise ($\vec{u}_t$).
Similarly,
the observation model $h(\cdot)$ describes current observations ($\vec{y}_t$)
as a function of the current state ($\vec{x}_t$) possibly subject to \emph{measurement} noise ($\vec{v}_t$).
For example,
in a tracking scenario,
the state vector may encode position and velocity,
$\vec{x}_t = [\vec{p}_t, \vec{v}_t]$
and the transition model $f(\cdot)$ may describe motion dynamics according to the laws of physics%
---the simplest case being inertial motion subject to random noise,
formally:
$\vec{p}_t = \vec{p}_{t-1} + \vec{v}_{t-1} \cdot \Delta t + \vec{u}_t$,
where $\Delta t$ is the time elapsed between $t-1$ and $t$.
In the same scenario,
observations may consist of noisy distance measurements from $M$ different sensors,
$\vec{y}_t = [d_{1,t}, d_{2,t}, \ldots, d_{M,t}]$,
where $d_{i,t}$ is the distance measured by sensor $i$ at time $t$,
and the observation model $h(\cdot)$ may relate these measurements to the underlying state
-- e.g., through geometric relationships --
accounting for sensor noise and biases.

\paragraph{The filtering problem}

Under these assumptions,
the filtering problem consists of recursively estimating the posterior probability distribution:
\begin{equation}\label{eq:posterior}
p(\vec{x}_t \mid \vec{y}_{1:t}),
\end{equation}
given a prior $p(\vec{x}_{t-1} \mid \vec{y}_{1:t-1})$
-- namely, the probability distribution of the previous system state w.r.t. all past observations --
and a new observation $\vec{y}_t$.

In linear-Gaussian settings,
the posterior from \cref{eq:posterior} admits a closed-form solution computed by the Kalman filter~\cite{Kalman1960}.
However,
many \ac{IoT} scenarios violate these assumptions due to
non-linear dynamics,
non-Gaussian noise,
multimodal beliefs,
or
abrupt regime changes.
In such cases,
Kalman-based approaches may
\begin{inlinelist}
    \item yield biased estimates,
    because the linear-Gaussian modelling would not capture the true system behaviour;
    \item diverge
    because the filter fails to track sudden changes in the state;
    or
    \item become computationally intractable.
\end{inlinelist}
These limitations motivate the use of more general Bayesian filtering techniques,
such as \ac{PF}.

\paragraph{\acf{PF}}

\ac{PF} addresses the general filtering problem by
approximating the posterior distribution through a finite set
-- of fixed cardinality $N$ --
of weighted samples
called ``particles''.
Hence,
\cref{eq:posterior} is approximated as:
\begin{equation}
p(\vec{x}_t \mid \vec{y}_{1:t}) \approx \sum_{i=1}^{N} w_{t,i} \cdot \delta(\vec{x}_t - \vec{x}_{t,i}),
\end{equation}
where $\{\vec{x}_{t,i}\}_{i=1}^N$ are particles
at time $t$,
$\{w_{t,i}\}_{i=1}^N$ are normalised weights,
and
$\delta(\cdot)$ is the Dirac delta function.

In practice,
such modelling allows computing the \ac{MMSE} estimate of the system state as the weighted average of the particles:
\begin{equation}
\hat{\vec{x}}_t = \sum_{i=1}^{N} w_{t,i} \cdot \vec{x}_{t,i}.
\end{equation}
In this way,
the whole estimation process reduces to maintaining and updating the set of particles and weights over time,
as new observations become available.

Each iteration of a \ac{PF} consists of three conceptual steps: prediction, weighting, and resampling.
During \emph{prediction},
particles are propagated according to a proposal distribution,
often derived from the state transition model.
Weights are then \emph{updated} based on the likelihood of the new observation.
Over time,
this process tends to concentrate probability mass on a small number of particles,
a phenomenon known as particle degeneracy.
To mitigate this effect,
\emph{resampling} redistributes particles by replicating those with higher weights and discarding those with negligible contribution.



\subsection{\acf{DPF}}
\label{sec:background-dpf}

The classical formulation of \ac{PF} assumes that
all observations are centrally available and processed by a single estimator.
This assumption is often incompatible with large-scale \ac{IoT} systems,
where sensing, computation, and communication resources are inherently distributed across multiple devices,
and where centralized data collection may be infeasible due to bandwidth, latency, energy, or robustness constraints.


In \ac{DPF},
the system state $\vec{x}_t$ remains conceptually global,
but \emph{observations} are taken from $K$ spatially distributed devices,
ranged through by $k$.
We denote by $\vec{y}_{t, k}$ the observation collected at time $t$ by device $k$.
There is no single observation model anymore,
but rather a set of \emph{local observation models},
$
\vec{y}_{t, k} = h_k(\vec{x}_t, \vec{v}_{t, k}),
$
where $h_k(\cdot)$ may vary across devices to account for heterogeneous sensing
capabilities, or viewpoints.
%

The goal of \ac{DPF} is to approximate the same posterior as in the centralised case,
$p(\vec{x}_t \mid \vec{y}_{1:t,1:K})$,
while relying only on local computations and limited inter-device communication.

\paragraph{Design space of \ac{DPF}}

Unlike centralised \ac{PF}, there is no unique way to distribute the filtering process.
Existing \ac{DPF} approaches explore a broad design space,
characterised by different choices regarding
\begin{inlinelist}
    \item how particles are represented and maintained across devices;
    \item what information is exchanged between devices,
    e.g., raw measurements, particle weights, sufficient statistics, or parametric approximations;
    \item how communication is structured,
    e.g., via neighbourhood exchanges, consensus protocols, or hierarchical aggregation;
    and
    \item where and how global consistency is enforced.
\end{inlinelist}

A first major distinction is between approaches that rely on a \emph{fusion centre}
and fully decentralised schemes.
Fusion-centre-based solutions delegate the final combination of information to a designated node,
which simplifies algorithm design but introduces a single point of failure and scalability bottlenecks.
Fully decentralised \ac{DPF} schemes,
instead,
aim to avoid any privileged node,
typically relying on peer-to-peer communication patterns.

Another common distinction concerns the role of particles.
Some approaches maintain identical particle sets across devices,
requiring synchronisation or consensus mechanisms to ensure consistency of particle weights.
Others allow each device to maintain its own local particle population,
exchanging information to progressively align local beliefs.
These choices impact communication overhead, 
robustness, and sensitivity to network dynamics.

\paragraph{Algorithmic challenges}

Distributing \ac{PF} introduces challenges that are absent in the centralized setting.
First,
\emph{communication constraints} limit how much information can be exchanged,
forcing approximations that may degrade estimation accuracy.
Second,
\emph{asynchronous execution} and \emph{time-varying network topologies} complicate the coordination of
prediction,
weighting,
and
resampling steps.
Third,
\emph{heterogeneity} in device capabilities and sensing models requires algorithms to gracefully handle
partial,
delayed,
or
unreliable information.

As a result,
state-of-the-art \ac{DPF} solutions typically embody flexibility through specialised algorithmic designs,
each tied to specific architectural and modelling assumptions.
While effective in their target scenarios,
such specialisations often limit adaptability to open and heterogeneous \ac{IoT} systems,
where network structure, device roles, and sensing modalities may change over time.

\paragraph{Towards flexible \ac{DPF} implementations}

From a system-level perspective,
\ac{DPF} can be interpreted as a \emph{distributed coordination} problem:
multiple devices collaboratively maintain and evolve a shared probabilistic representation of system state.
Such a perspective naturally raises questions about
programming abstractions,
coordination mechanisms,
and
adaptability,
beyond the design of individual algorithms.
These questions motivate the exploration of higher-level, declarative approaches to distributed estimation,
able to express \ac{DPF} logic independently of specific network structures or communication patterns,
as discussed in the following. 

\subsection{The \acl{FC} and \acl{AC}}
\label{sec:background-ac}


\emph{\acl{AC}}~\cite{DBLP:journals/computer/BealPV15} is a macro-programming~\cite{DBLP:journals/csur/Casadei23} and coordination paradigm
formally grounded in the \acl{FC}~\cite{DBLP:journals/jlap/ViroliBDACP19}
 that supports the design of the \emph{collective, self-organising behaviour}
 of large-scale, open, and dynamic networks of devices. 
Specifically,
 this is achieved through
 (i) a flexible execution model
   supporting ``continuous'' decentralised coordination,
 (ii) a functional programming model
   with field-based abstractions
   supporting global-level reasoning
    and modular composition of resilient collective behaviours.
These ingredients are detailed next.

%
%

\paragraph{System model and execution assumptions}

\ac{AC} assumes a system composed of many devices,
each endowed with local sensing, computation, and communication capabilities.
Devices interact by exchanging information with neighbouring devices,
according to an application-dependent notion of \emph{neighbourhood}
(e.g., spatial proximity or logical connectivity).
Execution proceeds in \emph{asynchronous ``perceive--compute--act'' rounds}:
each device repeatedly
\begin{inlinelist}
    \item \emph{perceives} its local context (given by sensor data and messages from neighbours),
    \item \emph{computes} local outputs,
    and
    \item \emph{acts} in its local context, e.g., by running movement actuations and disseminating locally-computed information to the neighbours.
\end{inlinelist}
No global clock, central coordinator, or reliable synchronisation is assumed.
%
%

\paragraph{Programming model: functions and computational fields}
In \ac{AC} there is a single \emph{macro-program}
  run by every device in their compute step.
The important aspect about the \emph{aggregate macro-programming model}
 is its ability to use
 neighbourhood communication
 within classical \emph{functions}~\cite{DBLP:journals/lmcs/AudritoCDV23}.
%
So, calling a function means such a function is executed by all the devices, round by round,
 with local outputs progressively changing
 also based on neighbour messages (prescribed by the function itself or its inputs).
This enables
 to capture reusable self-organisation patterns
 as functions,
 and to use \emph{functional composition}
 to build more complex self-organisation logic.

The fundamental abstraction underlying \ac{AC} is that of a
\emph{computational field},
i.e.,
a distributed data structure mapping each device
(also: the spatial locations where devices are situated)
to a value.
Conceptually,
a field is 
a mapping from $n$-dimensional space-time $\mathbb{R}^n \times \mathbb{R}_{\geq 0}$
to some value domain $\mathbb{Y}$---be it a scalar (e.g., temperature),
 a vector (e.g., velocity),
or a more complex object (e.g., a set of observations).
This abstraction admits a dual interpretation:
%
(i) \emph{local}: each device computes a value as part of its execution;
or
(ii) \emph{global}: the set of all values constitutes a field describing the system state as a whole.
This duality allows aggregate programs to be written as if manipulating global objects,
while being executed purely through local behaviours and interactions.
Functions can be then thought of
 as accepting and returning fields.

\paragraph{From primitives to building blocks}

Field calculi offer a small set of
 constructs 
 to build field-based computations:
\begin{inlinelist}
   \item observation of neighbouring values,
   \item stateful value transformation (round-by-round),
   and
   \item branching of computations over subsets of devices based on Boolean conditions.
\end{inlinelist}

Building on these primitive mechanisms, 
more complex patterns of distributed coordination can be expressed;
in particular,~\cite{DBLP:journals/tomacs/ViroliABDP18} identifies a core set of reusable self-organising \emph{building blocks}
providing:
\begin{inlinelist}
    \item information spreading and outward computation through a distributed gradient (cf~\cite{DBLP:conf/saso/WolfH07}.);
    \item information collection and inward aggregation through converge-cast;
    and
    \item distributed leader election~\cite{DBLP:conf/acsos/PianiniCV22}.
\end{inlinelist}
Crucially,
these patterns and any compositions thereof
are formally proven be self-stabilising~\cite{DBLP:journals/cacm/Dijkstra74}.

\paragraph{\acl{AC} in practice}
Practical usages of \ac{AC} are supported through \acp{dsl},
implementing variants of the \ac{FC}~\cite{DBLP:journals/jlap/ViroliBDACP19,DBLP:journals/tocl/AudritoVDPB19,DBLP:journals/lmcs/AudritoCDV23,DBLP:journals/jss/AudritoCDSV24}
that enable building layered libraries.
These \acp{dsl} can be found
both stand-alone~\cite{DBLP:conf/sac/PianiniVB15} and internal to mainstream
embedded in mainstream 
languages such as Scala~\cite{DBLP:journals/lmcs/AudritoCDV23},
C++~\cite{DBLP:journals/scp/AudritoT24}, and, more recently, Kotlin~\cite{cortecchia2024acsos}.
%
%
%
%
%
%
In this paper, we adopt such Kotlin-based \ac{dsl},
called \emph{Collektive}, as reference \ac{AC} language, 
detailing in \cref{sec:model} how it can be used to 
 implement \ac{DPF}.
%
%

\subsection{Related works}
\label{sec:related}


Distributed filtering has a long tradition in \acp{WSN} and multi-agent systems,
with \ac{DPF} methods surveyed and taxonomised in~\cite{HlinkaHD13}.
The survey categorises \ac{DPF} methods into families w.r.t.
how information is exchanged and where computation takes place, focusing on:
\begin{description}
    \item[consensus/gossip-based] schemes,
    where local filters are coupled through iterative agreement on sufficient statistics
    or likelihood\footnotemark{} surrogates~\cite{Olfati-Saber05,CattivelliS10};
    \item[likelihood\textsuperscript{\ref{note:likelihood}} fusion] approaches,
    where the key challenge is to approximate a joint likelihood\textsuperscript{\ref{note:likelihood}}
    (or its parameters)
    in a decentralised manner,
    enabling local particle updates that reflect global information~\cite{HlinkaSHDR12};
    \item[proposal adaptation and cooperative resampling] mechanisms,
    meant to mitigate weight degeneracy
    and
    improve efficiency under communication constraints~\cite{HlinkaHD14}.
\end{description}

\footnotetext{\label{note:likelihood}
  In the context of Bayesian filtering,
  the term ``likelihood'' commonly refers to the probability of the
  (possibly distributed)
  measurement vector ($\vec{y}_t$) conditioned on the system state ($\vec{x}_t$),
  i.e.\ $p(\vec{y}_t \mid \mathbf{x}_t)$.
}

Across these families,
solutions typically embody flexibility through specialised algorithmic designs,
but their implementation is still tightly coupled to specific architectural assumptions
(e.g., synchrony, connectivity, roles, message schedules),
which limits portability across heterogeneous and open \ac{IoT} deployments~\cite{HlinkaHD13}.
%
%
%
%
What is still missing is a unifying programming-model perspective that
\begin{inlinelist}
    \item treats \ac{DPF} and its components as  reusable, composable building blocks
    and
    \item cleanly separates filtering semantics from spatial coordination mechanisms,
\end{inlinelist}
so that the same estimator can adapt to changing
neighbourhoods,
densities,
and
connectivity.
Our work targets this gap by recasting \ac{DPF} as a field-based computation in \ac{AC},
enabling flexible orchestration of particle evolution
and
information fusion through aggregate operators,
rather than through architecture-specific protocols.

\section{Field-based \acp{DPF}}
\label{sec:model}

As discussed in~\Cref{sec:background}, 
 existing \ac{DPF} solutions can be broadly classified into three architectural families: 
\begin{enumerate*}[label=(\roman*)]
 \item consensus-based approaches; 
 \item fusion-center-based schemes; and 
 \item leader-agent-based formulations~\cite{HlinkaHD13}.
\end{enumerate*}
While these approaches explore different trade-offs between accuracy, communication cost, and robustness, 
 they are typically grounded in strong assumptions on network topology, communication structure, and node roles, 
 which limits their adaptability in open and dynamic IoT deployments.

In this section, 
 we show how \ac{AC} enables a unified modelling of DPF that abstracts over such architectural commitments. 
By expressing estimation and coordination as field-based computations, 
 AC decouples the filtering logic from the underlying network structure, 
 yielding designs 
 that remain valid under dynamic changes in density and topology. 

\subsection{Architectural Abstraction and Coordination}
\label{sec:architecture}

From an architectural standpoint, 
 \ac{AC} naturally subsumes the main \ac{DPF} families within a single programming model. 
Fusion-center-based solutions can be expressed without statically designating a central node, 
  by realising the fusion role as a dynamically elected leader. 
Unlike classical leader-agent approaches, 
 where leadership may change at every filtering step, \ac{AC} supports \emph{self-stabilising leadership}~\cite{DBLP:conf/acsos/PianiniCV22,DBLP:journals/automatica/MoADB22},
 in which the leader persists and is replaced only in response to failures or significant topological changes. 
This results in fusion-like behaviour combined with fault tolerance, self-healing, 
 and the absence of single points of failure.

More generally, \ac{AC} enables 
flexible interpolation between fully decentralised and centrally coordinated DPF schemes,
 without requiring algorithmic redesign. 
Architectural choices thus become configuration aspects of the aggregate program 
 rather than intrinsic properties of the filtering algorithm.

\subsection{Information Exchange as Field Computation}
\label{sec:information-exchange}

\ac{AC} also provides a principled abstraction for modelling what information is exchanged among nodes during DPF execution. 
In the literature, 
 information sharing typically involves either
\begin{inlinelist}
    \item particle sets,
    \item parametric approximations of the posterior (e.g., Gaussian or mixture-of-Gaussians statistics), or
    \item global likelihood surrogates~\cite{HlinkaHD13}.
\end{inlinelist}
All these strategies can be naturally represented as computational fields, 
 locally produced and selectively propagated or aggregated across the network.

Beyond these established approaches, 
 \ac{AC} enables an additional design option based on the aggregation of raw local measurements. 
Let each of the $K$ sensors acquire a local observation at time $t$ according to
\begin{equation}
y_{k,t} = h_k(x_t, v_{k,t}), \qquad k = 1, \dots, K.
\end{equation}
Using aggregate operators, 
 local measurements can be combined within neighbourhoods or dynamically defined regions $\mathcal{N}$,
 yielding an aggregated measurement function:
 \begin{equation}
     \hat{y}_t = H_{\mathcal{N}(k)}\big( \{ h_k(x_t, v_{k,t}) \}_{k \in \mathcal{N}} \big),
\end{equation}
which can be used directly in the particle weighting step.

This form is equivalent to the construction of a distributed sensor, 
 where multiple devices jointly contribute to a single, higher-quality observation.
Moving from local to distributed sensing is especially beneficial in scenarios 
 with limited, noisy, or fragile 
 sensors, 
 where aggregating raw measurements can significantly improve the quality of the likelihood estimation. 
To the best of our knowledge, 
 existing \ac{DPF} approaches like ~\cite{rosencrantz2003uai,coates2004ipsn, figueredo2020spl}
  mainly focus on sharing 
  particles, beliefs, or likelihood approximations, 
 while the explicit construction of aggregated measurement functions by raw measurement sharing has not been systematically explored.

\subsection{Spatially Adaptive Computation \&  Partitioned Estimation}

A further advantage of integrating \ac{DPF} with \ac{AC} lies in the ability 
 to dynamically adapt the spatial extent of computation based 
 on the estimated relevance of sensing nodes with respect to the system state.

In many distributed estimation scenarios, 
 and in particular in target tracking, 
 only a subset of sensors provides informative observations at any given time. 
Sensors that are far from the object of interest typically produce highly noisy, weak, or even absent measurements, 
 contributing marginally to estimation accuracy while still incurring computational and communication costs. 
Within \ac{AC}, 
 this situation can be naturally addressed by embedding the restriction directly into the aggregate program, 
 using domain separation (distributed branching) constructs that dynamically partition the network based on local observability conditions
 (e.g., proximity or measurement quality with respect to the target).
 
As a result, 
 particle filtering can be selectively activated only on nodes that are sufficiently close to the estimated object location, 
 while the remaining nodes remain computationally inactive or execute a lightweight monitoring role.

This form of spatial gating of computation is particularly easy to express in AC through field-based conditionals, 
 and does not require explicit role assignment or protocol changes. 
 
Importantly, 
 as the estimated state of the object evolves over time, 
 the region of active computation smoothly follows it, 
 automatically activating newly relevant sensors and deactivating those that become uninformative. 
This mechanism may yield improved energy efficiency and scalability,
 while simultaneously reducing the impact of low-quality measurements on the estimation process.

Beyond selective activation, 
 \ac{AC} also supports the construction of partitioned network architectures through the 
 Self-Organising Coordination Regions (SCR) pattern~\cite{DBLP:journals/fgcs/PianiniCVN21},
 whereby the sensor network is dynamically partitioned into spatial coordination regions inducing a Voronoi-like tessellation 
 of devices based on their positions.
Within each partition, 
 a representative device (i.e., the leader) is elected to coordinate local estimation activities, 
 aggregate information, or interface with higher-level coordination layers. 
Such architectures are especially beneficial in settings characterised by severely constrained communication capabilities, 
 or when multiple objects are simultaneously present in the environment.
In these scenarios, 
 each partition can focus on estimating the state of the objects located within its spatial domain, 
 while ignoring distant objects whose observations would be weak or irrelevant. 
Global consistency can then be recovered through limited coordination among partition leaders only, 
 significantly reducing communication overhead. 
Crucially, 
 these partitions and leadership roles are not statically defined: 
 they emerge from the aggregate program and continuously adapt to changes in node density, topology, or object distribution.

Moreover, 
 while partitions may initially be constructed solely based on spatial proximity, 
  AC enables their refinement through additional metrics, such as signal quality, uncertainty, or task-specific relevance. 
This can be achieved by exploiting the notion of space-fluidity~\cite{DBLP:journals/lmcs/CasadeiMPVZ23}.
As a result, partitions may expand, contract, or deform over time, 
 better aligning the computational structure with the evolving estimation needs.

All the above mechanisms are realised within AC in a self-organising and self-healing manner. 
Node failures, mobility, or the disappearance of elected leaders do not require changes to the aggregate specification: 
 the system automatically reconfigures itself, re-electing leaders, reshaping partitions, 
 and re-routing information flows as needed. 
This further reinforces the suitability of AC as an abstraction layer 
 for engineering robust and adaptive DPF solutions in open IoT environments.

%

\section{Exemplary case: target tracking}
\label{sec:use-case}

In this section,
we show a possible application of \ac{AC} as a technique to implement \acp{DPF},
focusing on a common problem in the \ac{IoT} domain:
target tracking through a network of spatially distributed sensors.

\subsection{Problem description}
\label{sec:model-design}

We consider a target-tracking problem in a bidimensional Euclidean space. 
We assume such space to be populated by a set of spatially distributed sensing devices,
forming a static sensor network,
$S = \{s_1, \ldots, s_k\}$ deployed in fixed positions,
tracking a single moving target $T$ whose trajectory is unknown.
Each sensor $s_i$ is capable of minimal computation 
 and communication within a limited neighbourhood $\mathcal{N}(s_i)$.

Also,
 we assume the sensing devices can perceive the target 
 by the radio signal 
 emitted by the moving object.
Each sensor $s_i$ is located at a known position $\mathbf{r}_i = (x_i, y_i)$,
 while the position of the moving target $T$ is denoted by $\mathbf{m} = (x, y)$.
The distance between the moving target $T$ and sensor $s_i$ is defined
 as $d_i = \| \mathbf{m} - \mathbf{r}_i \|$.
The perceived signal power at sensor $s_i$ is modeled using a log-distance path-loss model:
\begin{equation}
P_i = P_0 - 10 n \log_{10}\left(\frac{d_i}{d_0}\right) + \varepsilon_i,
\end{equation}
where $P_0$ is the received power at a reference distance $d_0$, $n$ is the path-loss exponent,
 and $\varepsilon_i \sim \mathcal{N}(0, \sigma_p^2)$ models measurement noise.

Finally, 
 each sensor runs an aggregate program. 
Execution of devices proceeds in semi-synchronous rounds:
 each device executes rounds at approximately the same rate,
 but with bounded drift between any two subsequent rounds.
This matches the standard execution model assumed by \ac{AC} and \ac{FC}.
This is also 
 why we index time with discrete steps $t \in \mathbb{N}$,
 despite the underlying system evolves in continuous time.
Also, notice that the notion of \emph{round} here refers to the local computation cycle of each device,
 so the same round index $t$ may correspond to slightly different real times across devices.

\subsection{Aggregate Measurement Functions}
As discussed in Section~\Cref{sec:information-exchange}, 
 a key advantage of leveraging \ac{AC} for \ac{DPF} lies in the ability
 to construct aggregate measurement functions from local observations shared among neighbouring devices.
Rather than treating each sensor as an isolated information source, \ac{AC} enables 
 the sensing infrastructure to be interpreted as a computational field of measurements, 
 where local perceptions are continuously combined through neighbourhood interactions.

In the considered target-tracking scenario, each sensor $s_i$ acquires at time t 
 a local measure $y_i^t$ of the position of the target $T$. 
Through neighbourhood communication, 
 each sensor gains access to the observations collected by its neighbours, 
 and can locally compute an aggregate measurement function that refines
 the estimation of the target position. 
Notice that as the number of neighbouring sensors increases, 
 the aggregate estimate becomes more accurate due to the availability
 of a richer set of local observations.
Such aggregation can take simple forms, 
 such as weighted averages of neighbouring measurements, 
 or more sophisticated functions that account for measurement quality, 
 for instance by weighting observations according to the estimated distance 
 from the target or the expected noise level.
In this work, 
 we consider an indirect form of measurement aggregation, 
 where observations from neighbouring sensors are treated as conditionally independent 
 and combined by multiplying their local likelihoods.

Importantly, 
 each sensor maintains its own local particle filter and its own set of particles. 
Aggregation operates solely at the level of measurements, 
 without requiring the exchange of particles among neighbours. 
This design choice significantly reduces communication overhead, 
 which would otherwise be dominated by the transmission of particle sets, 
 while still allowing each local filter to incorporate richer and more informative observations 
 through neighbourhood-level aggregation.

\subsection{Fusion Center as an elected leader}

As discussed in~\Cref{sec:architecture}, 
 a second advantage of adopting \ac{AC} for \ac{DPF} is the ability 
 to realise fusion-center-based architectures through dynamic leader election. 
Rather than selecting a fusion center a priori, 
 AC allows the fusion role to be assigned at runtime based on a chosen metric, 
 such as network centrality or other application-specific criteria.

This approach provides two main benefits. 
First, 
 the fusion center is not statically bound to a specific device, 
 but can adapt to changes in network topology or deployment conditions. 
Second, 
 in the presence of node failures, 
 the system automatically elects a new leader in a self-healing and self-organising manner, 
 without requiring external intervention, allowing the tracking process to continue over time.

In this setting, 
 the system still exposes a computational field of local measurements $y_i^t$, 
 which are progressively aggregated toward the elected leader through a converge-cast operator. 
The leader node thus acts as a fusion center, 
 collecting measurements from the network and executing a particle filter based 
 on the received information.

Notice that measurements from sensors 
 at different hop distances reach 
 the fusion center with different delays: 
 data from neighbours arrive immediately, 
 while information from farther nodes experiences multi-hop propagation delays. 
Consequently, 
 updates at the fusion center are temporally staggered 
 and reflect the underlying network topology.

\begin{figure*}
  \centering
  \includegraphics[width=\textwidth]{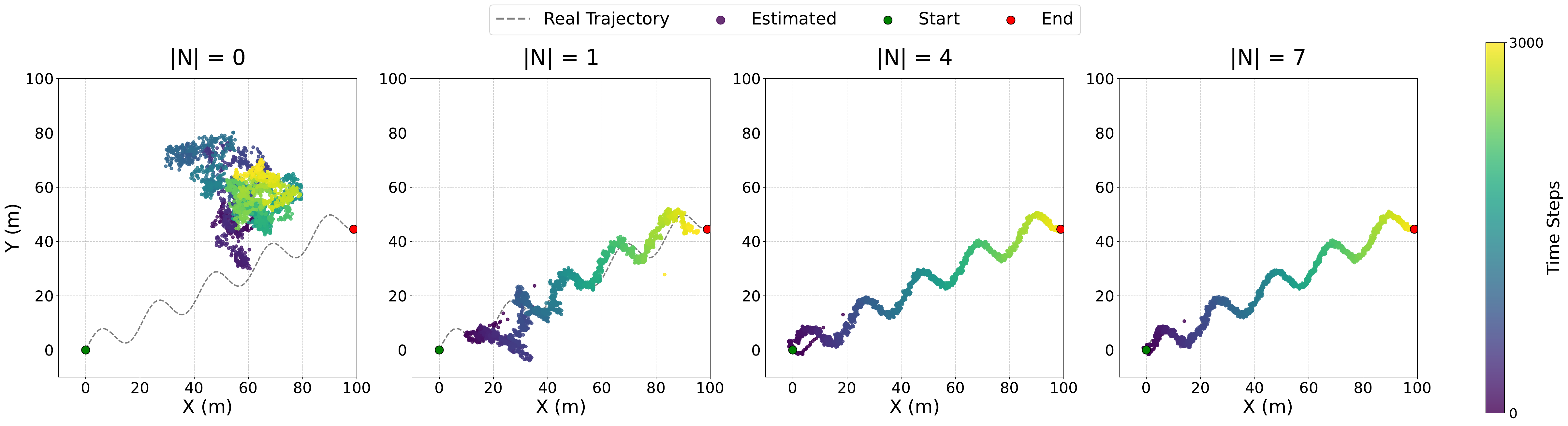}
  \caption{
    Estimated trajectories for different numbers of sensors. With few neighbors (0-1), 
    the filters fail to converge due to limited information; 
    increasing the neighbors count allows exploiting aggregated measurements, 
    improving the estimation and enabling convergence toward the real trajectory.
  }
  \label{fig:exp1-trajectories}
\end{figure*}

\begin{figure}
  \centering
  \includegraphics[width=0.9\columnwidth]{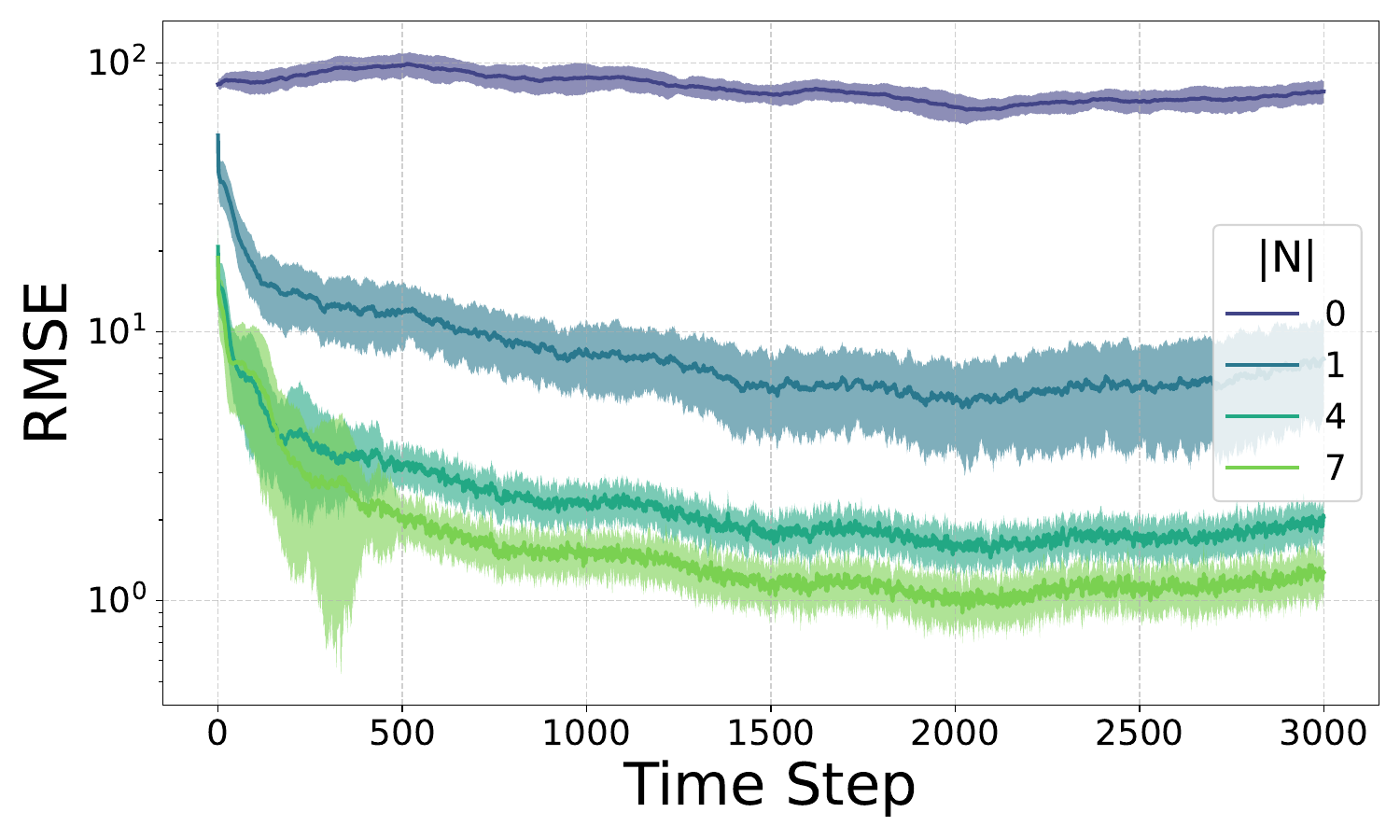}
  \caption{
    Average \acs{RMSE} over 100 random seeds,
     with shaded areas indicating variance.}
  \label{fig:exp1-rmse}
\end{figure}

\begin{figure}
  \centering
  \includegraphics[width=0.8\columnwidth]{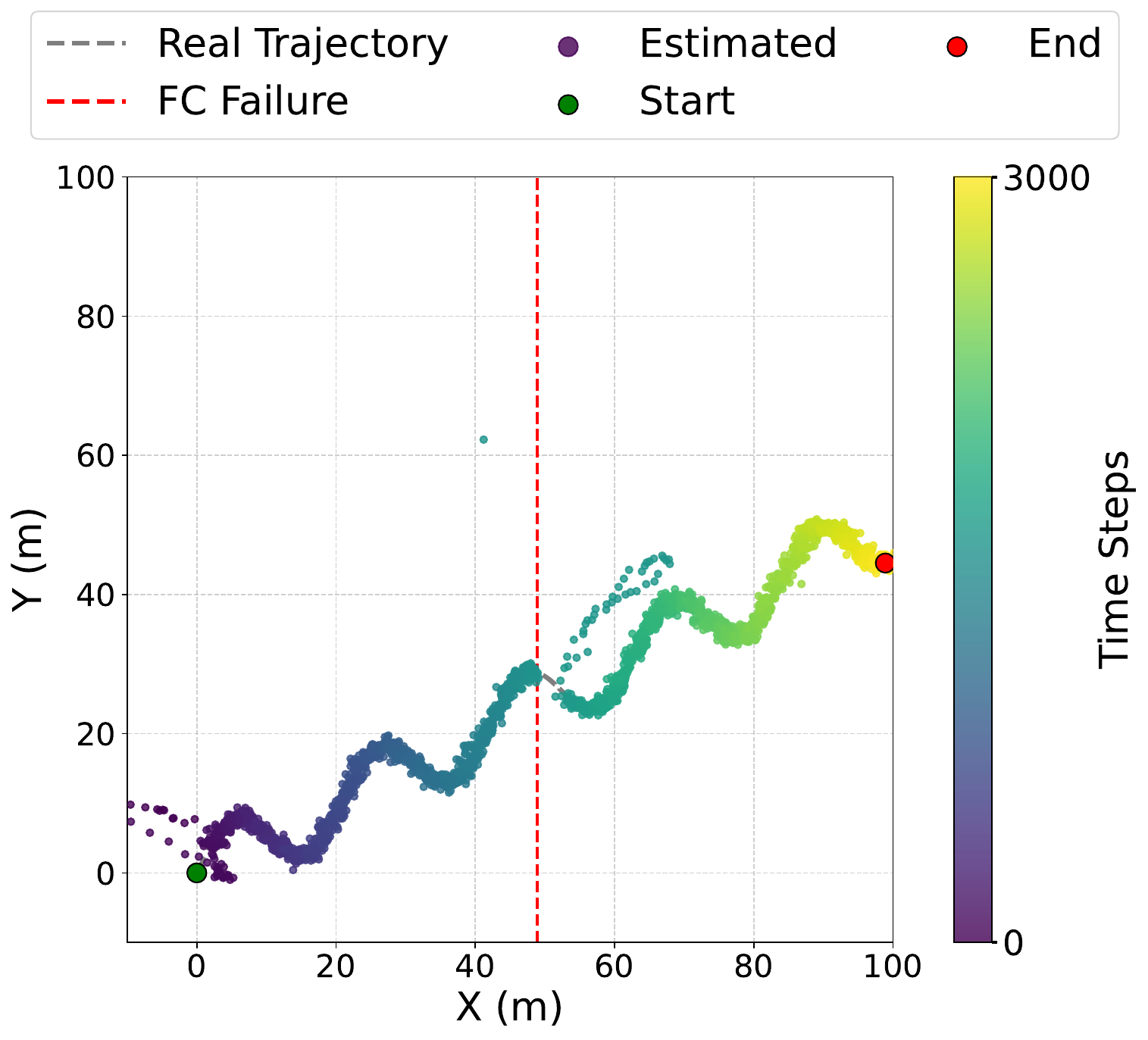}
  \caption{
    Trajectories estimated via the leader-based fusion center approach.
    The plot shows the estimated trajectory before and after the leader failure at time step $1500$.
    A brief transient phase is observed 
    initially and around the disconnection event,
    which corresponds to the leader election process.
  }
  \label{fig:exp2-trajectories}
\end{figure}

\section{Experimental Evaluation}
\label{sec:experiments}

\subsection{Experimental Setup}\label{subsec:experimental-setup}

To validate the approach described in the previous sections,
 we conducted a simulation-based evaluation.
All the experiments are publicly available 
 on GitHub\footnote{\url{https://github.com/domm99/experiments-ac-based-distributed-particle-filtering}}
 under a permissive license for reproducibility.
Simulations were developed using the Alchemist simulator~\cite{Pianini_2013},
 which provides an environment for simulating distributed, dynamic systems and \ac{AC} programs.
The program logic for the sensors' behaviour and program evaluation was implemented in Kotlin,
 leveraging the Collektive framework for \ac{AC} constructs~\cite{cortecchia2024acsos}.

We designed two main experiments to validate the approaches described in~\Cref{sec:use-case}:
\begin{inlinelist}
    \item an experiment focusing on evaluating an aggregated measurement function through local measurement sharing among neighbouring sensors;
    \item an experiment focusing on evaluating the leader-based fusion centre approach for reconstructing the target trajectory,
    even under dynamic leader changes.
\end{inlinelist}

The experiments consist of a network of $25$ sensors deployed on a two-dimensional spatial grid,
 with slight random perturbations of the nominal grid positions.
In the environment,
 there is a single moving object following a predefined
 non-linear trajectory $\gamma$:
\begin{equation}
  \gamma:\mathbb{R}\to\mathbb{R}^2,\qquad
  \gamma(t)=\left(vt,\ \frac{vt+5\sin(0.3\,vt)}{2}\right).
\end{equation}
where $v = 0.05\nicefrac{m}{s}$ is a constant speed.
The moving object emits signals that can be perceived by the sensors: 
 the further the object is from a sensor, 
 the weaker the perceived signal, 
 and the noisier the measurement becomes.

The sensors and object states are defined by their planar position $(x,y)$,
 and the moving object state is defined by its velocity components along the two axes.
Sensors do not move,
and run computations at a fixed 1Hz rate, with no syncronisation between them.
The neighbourhood size $|N|$ is a free variable of the experiment.
Each configuration executes for
$3000$ simulated seconds.
Every experiment was 
repeated $100$ times
with different random seeds
to account for the stochasticity.

The goal of the sensors is to cooperatively track the moving object
 by estimating its position over time via \ac{DPF}. 

\subsection{Discussion}\label{subsec:discussion}

\Cref{fig:exp1-trajectories} shows the estimated trajectories for the first experiment, 
 in which each sensor runs its own particle filter while constructing 
 an aggregated measurement function by sharing local measurements with its neighbours.
The figure reports the performance obtained for different neighbourhood sizes ($|N|$).
When no neighbours are available, 
 the filters fail to converge due to the limited amount of information.
As neighbouring sensors increase, 
 aggregated measurements improve accuracy and convergence.
A single neighbour is sufficient to improve the estimation
significantly over time.
With $|N|=4$,
 a brief initial instability is observed, 
 followed by rapid convergence.
With $|N|=7$ neighbours,
 the estimation is both accurate and stable over time,
 highlighting the benefits of aggregating information from multiple sensors.
The same trend is observed in \Cref{fig:exp1-rmse},
 which reports the average \ac{RMSE} over time.
With few neighbours, 
 the estimation error remains high and does not decrease,
 whereas exploiting information from a larger neighbourhood
 significantly reduces the error and improves long-term stability.

\Cref{fig:exp2-trajectories} shows the estimated trajectories for the second experiment,
 where a leader-based fusion center is used to reconstruct the target trajectory.
The chart shows the estimated trajectory before and after the leader failure at time step $1500$.
We observe a brief transient phase associated with leader election at the beginning of the simulation, when tracking starts,
 and a second transient triggered by the failure of the current leader and the subsequent election of a new one.
Once the new leader is elected, after a brief transient period,
 it is able to take over the fusion center role and continue tracking the moving object.
This shows the resilience of the leader-based fusion center approach,
 as it can adapt to dynamic changes in leadership and maintain accurate tracking performance.

\section{Conclusion and Future Work}
\label{sec:conclusion}
In this paper,
  we propose a field-based formulation of \ac{DPF} grounded in \ac{AC}, 
  decoupling the estimation logic from architectural and coordination concerns. 
By expressing sensing, 
 data sharing, 
 and fusion mechanisms as computational fields, 
 our approach enables the systematic derivation of different DPF configurations within a unified programming model. 
The experimental evaluation on a target-tracking scenario shows
 that both neighbourhood-level aggregated measurement functions 
 and leader-based fusion centres can be effectively realised in \ac{AC},
 yielding improved estimation accuracy and robustness while preserving decentralised coordination.

Future work will focus on experimentally exploring more 
 design dimensions discussed in~\Cref{sec:model} but not evaluated in this paper. 
In particular, 
 we plan to investigate spatially adaptive computation strategies, 
 including dynamic restriction of filtering activities to relevant regions 
 of the network and partitioned estimation through self-organising coordination regions. 
Finally, 
 extending the evaluation to multiple targets and heterogeneous sensing modalities is a  
 natural next step to further assess the flexibility of the proposed framework.
 
\section*{Acknowledgments}
This work contributes to the research agenda of the \emph{Italian Science Fund (FIS3)} Starting Grant project \emph{FoMaSE -- Foundations for Macro-programming-based Software Engineering} (Grant No. FIS-2024-00174, CUP J53C25002170001).

\bibliographystyle{IEEEtran}
\bibliography{bibliography}

\end{document}